
\input harvmac
\Title{HUTP-91/A059}{Topological Mirrors and Quantum Rings}
\vglue 1cm
\centerline{{Cumrun Vafa}\foot{Talk presented at MSRI conference on Mirror
Symmetry, March 1991, Berkeley.}}
\centerline{Lyman Laboratory of Physics}
\centerline{Harvard University}
\centerline{Cambridge, Ma 02138, USA}
\vglue 2cm
Aspects of duality and mirror symmetry
in string theory are discussed.  We emphasize, through examples,
the importance of loop spaces for a deeper understanding of the geometrical
origin of dualities in string theory.
Moreover we show that mirror symmetry can
be reformulated in very simple terms as the
statement of equivalence of two classes of topological theories:
Topological sigma models and topological Landau-Ginzburg models.
Some suggestions are made for generalization of the notion of mirror symmetry.

\Date{11/91}

\newsec{Introduction}

One of the most fascinating aspects of string theory is the way
it modifies our intuition of classical geometry.  It modifies
it in ways which in some sense makes the classical geometry
{\it more symmetrical}, and thus, in a sense simpler. This
is probably most manifest in the principle of duality
in string theory, which states that two
classically inequivalent geometries (target spaces for strings)
can nevertheless be identical
from the string point of view.  The aim of this paper is to
develop this notion emphasizing the basic physical reasons for believing in
its universal existence.  My presentation is written
with the mathematically oriented reader in mind and even though I will not
be fully rigorous I hope that the main ideas
are more or less clear to mathematicians.

I will first discuss some general aspects of
Hilbert space of strings propagating in a target space in a geometrical
way and discuss the notion of duality in this set up (section 2).
Then I give some simple examples of this duality for
bosonic strings (section 3).
In section 4, I will discuss aspects of fermionic (super-) string vacua
highlighting aspects which are relevant for mirror symmetries.
As we will see an important ingredient in this setup is the notion
of {\it quantum cohomology ring} of Kahler manifolds which is a
deformation of the ordinary cohomology ring.
In section 5 the relation between singularity theory and solutions
of superstrings is discussed.  This turns out
to be a convenient bridge between target space interpretation and
abstract conformal field theory definition of string theory.  In
section 6 the topological formulation of mirror symmetry is discussed.
This turns out to be a very effective language to describe mirror symmetry.
In this setup,
mirror symmetry is stated as the equivalence of two seemingly inequivalent
topological theories.
This topological formulation has the advantage
of simplifying the conformal theory to a much simpler
theory which is the relevant piece needed for
the discussion of mirror symmetry.
Finally in section 7 I discuss some puzzles for mirror symmetry
and their potential resolutions.  I also discuss some potential
generalizations of mirror symmetries and some possible connections
with quantum groups and Donaldson theory.

\newsec{String Hilbert space}
In this section we discuss the basic structure of string vacua
which involves the Hilbert space and operatorial formulation
of the theory (this aspect is discussed much more extensively in
the talks of Friedan in this conference;
for a mathematical introduction see \ref\Seg{G.
Segal, {\it Conformal Field Theory}, Oxford preprint; and lecture
at the IAMP Congress, Swansea, July, 1988.}).
Consider a closed string (one dimensional parametrized circle) sitting in a
Riemannian manifold
$M$.  The space of all such configurations is given by the (parametrized) loop
space of $M$ which we denote by ${\cal L} M$.  The geometrical questions
that arise in string theory basically correspond to probing the geometry
of ${\cal L} M$.  The Hilbert space of {\it bosonic strings} is
an `appropriate' category of function space on ${\cal L} M$, which
we denote by
$${\cal H}_{bosonic}=\Phi ({\cal L}M)$$
with norm inherited from the metric on $M$.  The Hilbert space of {
\it fermionic or superstrings}
is the space of semi-infinite forms on ${\cal L}M$:
$${\cal H}_{fermionic}=\Lambda^\infty ({\cal L}M)$$

In addition to this Hilbert space, there is a more or less
canonical one to one correspondence between the states $|v\rangle$
in the Hilbert space and
some `special' operators $O_v$ acting on the Hilbert space.
Roughly speaking, these
operators are characterized by the fact that they are `invariant'
under reparametrizations of the string and that when they act
on a special state $|0\rangle$ (the vacuum state) in the Hilbert space, they
give the corresponding state $(O_v|0\rangle =|v\rangle)$.  These form
a complete operator product algebra, in the sense that the product
of any two of these operator is another such operator.  Choosing
a basis, we have
$$O_i O_j=\sum_k C_{ij}^kO_k$$
where the sum over $k$ is generically an infinite sum.

A convenient method of computing $C_{ij}^k$ is as follows:
In string theory to find the amplitude of how a number of loops
$l_i\in {\cal L}M$  ends up changing to the loops
${\tilde l}_j \in {\cal L}M$ we have to sum over all interpolating surfaces
$\Sigma$ immersed in $M$, $f(\Sigma ) \subset M$ whose boundary is
$$\partial  f((\Sigma )) =\bigoplus l_i -\bigoplus {\tilde l}_j$$
weighed by $exp(-E)$ where $E$ is the energy functional of
the surface immersed in $M$
(a natural extension of this applies to fermionic strings).
We can choose a `basis' for our Hilbert space of
delta functions corresponding to fixed loops in the manifold.
The above prescription then gives a way to compute the amplitude
that two of these basis elements ends up with the third one.  This
can be extended to the full Hilbert space by multi-linearity of the
amplitude.  The amplitude thus computed for the two string state
$|i\rangle $ and $|j \rangle$ to end up with the third one $|k\rangle$
can be obtained by integrating the `wave function' of these states
against the basic amplitude with the delta functions.  The resulting answer
is in fact the same as $C_{ij}^k$.

 There are consistency conditions that
Hilbert space and
these coefficients need to satisfy for a consistent theory (following
from the associativity of the operator products and modular invariance
of string amplitudes).  Once we are given such a structure, we can
forget about $M$ altogether and talk about the `string vacuum',
meaning this abstract Hilbert space with some canonical set
of operators satisfying some `nice' operator product properties.  Let us denote
such a structure by $S$ and call it a {\it string vacuum}.
Then two string vacua are equivalent, or isomorphic, if
there is an isomorphism between the corresponding Hilbert spaces
and the operators.  Now it may happen that strings on two different manifolds
 $M_1$ and $M_2$ give rise to isomorphic
string vacua
$$M_1\not= M_2 \quad but \quad S(M_1)= S(M_2).$$
In other words {\it the map from manifolds to string vacua
may be many to one}.  In such a case we call the manifolds $M_1$ and
$M_2$ {\it dual or mirror pairs}.  Actually the choice of the terminology
is unfortunate, as it may happen that more than two manifolds may
give rise to the same string vacuum.  One could also ask the reverse
question:  Does every string vacuum come from a manifold, i.e.,
is this map onto?  The answer seems to be no (see for example \ref\asym{
K.S. Narain, M.H. Sarmadi and C. Vafa, Nucl. Phys. B288 (1987) 551\semi
J. Harvey, G. Moore and C. Vafa, Nucl. Phys B304 (1988) 269.}).

The existence of mirror symmetry is thus simply the statement of
the existence of different geometrical ways to realize a string vacuum.
We can use any representation we please.  In such
cases, if we try to study some aspects of the string vacuum we
can choose any realization and may thus end up
equating a `hard' geometrical computation
in one representation to an `easy' one in another realization.
In this lies the power of mirror symmetry transforming a hard
problem to an easy one.
In the next section we give some examples of mirror pairs in the context
of bosonic strings.

\newsec{Examples of Bosonic Mirrors}
In this section we consider examples of mirror manifolds which
lead to the same string vacuum for the bosonic strings.  We
will give two classes of examples:  In one class the mirror Riemannian
manifolds are topologically the same but geometrically distinct,
and in the second class the mirror manifolds are even topologically
distinct.

Let $M_1$ be the $d$ dimensional torus identified (as a Riemannian
manifold) with
$$M_1={E^d\over \Gamma}$$
where $\Gamma$ is a $d$ dimensional discrete lattice group acting
by isometry on flat Euclidean space $E^d$.  Let us consider the Hilbert space
of strings on $M_1$ which is related to the function space on ${\cal
L} M_1$.  First note that ${\cal L}M_1$ naturally splits to infinitely many
components, corresponding to each element of $\Gamma$ which can be
identified with $H_1(M_1,Z)$.  Moreover, the function space on each
component splits to the functions of the center of strings
which is isomorphic to ordinary function space on $M_1$, and functions
of oscillations of loops (which is universal and independent of $\Gamma$).
The function space on $M_1$ is canonically isomorphic to $\Gamma^*$, the
dual lattice to $\Gamma$ using Fourier transform. So the dependence
of Hilbert space of strings on $\Gamma$, appears as a choice of
loop component (an element of $\Gamma$) and the Fourier component
of functions of center of string (an element of $\Gamma^*$), i.e.,
the dependence  comes through a choice of element of
$$\Gamma +\Gamma^*$$
This implies that if we consider the second manifold $M_2$
$$M_2={E^d\over \Gamma^*}$$
Then the Hilbert spaces of strings based on $M_1$ and $M_2$ are
isomorphic, both depending on the {\it self dual}
lattice $\Gamma+\Gamma^*$.  This turns out to extend to the full
string vacuum structure, i.e., to the operators and their products.
So $M_1$ and $M_2$ are mirror pairs.
  In physical terms this implies
that there is no physical experiment one can do in string theory
to distinguish strings on $M_1$ from strings on $M_2$.
This means
in particular that the notion of `length' is not a universally
invariant way to decide if two manifolds are different as far
as strings are concerned.
  This simple example
illustrates the basic structure of duality or mirror symmetry in bosonic
strings. This in fact was the first example of mirror symmetry discovered
in string theory \ref\dual{K. Kikkawa and M. Yamasaki, Phys. Lett. B149
(1984) 357\semi N. Sakai and I. Senda, Prog. Theor. Phys. 75 (1984) 692.}.
 The rest of the examples are just extensions of this
to more intricate cases.

For our second class of example we consider a simply laced compact Lie group
$G$.  Let $H$ denote its Cartan torus.  Consider an element $g\in G$ of
finite order which belongs to the normalizer of $H$ (i.e., it acts
as a Weyl transformation on $H$). This means that
$$H\rightarrow gHg^{-1}$$
Let us denote the cyclic group generated by this transformation $\Lambda_1$
(we take $g$ to act non-trivially on $H$).
Choose an element  $h\in H$ conjugate to $g \in G$.  Consider the action
$$H\rightarrow h\ H$$
and denote the group action generated by this cyclic group $\Lambda_2$.
Consider taking the quotients of $H$ by these two different group actions:
$$M_1={H\over \Lambda_1}\ \qquad \ M_2={H\over \Lambda_2}$$
These two spaces are completely different.  In fact $M_1$ is
not even a manifold, but an orbifold, as $g$ acts by fixed points on $H$,
but $M_2$ is simply another torus, as $h$ simply generates translations on $H$.
It turns out that (bosonic) strings
propagating on $M_1$ and $M_2$ are equivalent \ref\orb{
L. Dixon, J. Harvey, C. Vafa and E. Witten, Nucl.Phys. B274 (1986) 285
\semi J. Lepowski, Proc. of the Nat. Acad. of Sci. 82 (1985) 8295.}.
It is somewhat surprising that $M_1$ which is not even a manifold
behaves very much like the smooth manifold $M_2$ as far as strings are
concerned.  This means, in the mathematical sense (as is also
seen in examples for superstrings \orb \ref\Ro{S.S. Roan,
Int. Jour. of Math., v.1 (1990) 211.})
that loop space of an orbifold is a far better behaved object
than the orbifold itself and in a sense provides a kind of universal space
for resolution of orbifold singularity.

It should also be clear from the
above examples that we can construct examples
where three (or more) inequivalent Riemannian manifolds lead to the
same string vacuum.

\newsec{Superstring Vacua and Quantum Cohomology Rings}
Most of our discussion up to now has been on bosonic strings.
This is the case in which the Hilbert space is roughly speaking the
function space on the loop space of manifold. However
fermionic string is the physically (and mathematically)
more interesting case.
This is the case corresponding to the Hilbert space of semi-infinite
forms on the loop space.  In most applications one considers
target spaces which are Kahler manifolds.  In this case
the Hilbert space and the operators acting on it naturally
admit $Z\oplus Z$ grading, corresponding to the (holomorphic,
anti-holomorphic) degree of the differential forms.  Let ${\cal O}$
denote the space of physical operators.  Then we have the decomposition
according to the degrees of the forms:
$${\cal O}=\bigoplus_{p,q \in Z}{\cal O}_{p,q}$$
Naturally under operator products the degrees add, as expected.
Note that since we are dealing with semi-infinite differential forms,
the degree of operators runs from $-\infty$ to $+\infty$.  This is an important
difference with respect to the differential forms on the ordinary manifolds
where the degree of differential forms is positive.
As we shall see later this
is one of the main reasons for the prediction of mirror
symmetry in the fermionic strings.  There is an anti-unitary involution
which implies that $O_{p,q}$ is the conjugate of $O_{-p,-q}$.
The existence of this anti-unitary involution is the statement of
$CPT$ invariance of the theory.  In the language of forms, since
we are dealing with semi-infinite forms, it is roughly the statement that
operation of `adding' and `subtracting' forms are conjugate operations.
This turns out to be an important piece of physics in the story of mirror
symmetry.

Since the manifold $M$ is naturally embedded in ${\cal L}M$, one expects
that at least the differential forms on $M$ are related to a subset
of those on ${\cal L}M$ and in particular the cohomology ring
of $M$ should correspond to some closed operator algebra
(modulo addition of cohomologically trivial elements) of
operators acting on the fermionic Hilbert space.  Let $d$ denote
the complex dimensions of $M$.  Then we expect that there exist
a special set of operators $A_{\alpha} \in {\cal O}_{p,q}$
with $0\leq p,q\leq d$, such that the operator algebra of
$A_{\alpha}$ correspond to the cohomology ring of $M$.
  This expectation
turns our to be correct and we denote this subsector
of the operators by $H^{*,*}$.  In fact more is
true \ref\lvw{W. Lerche, C. Vafa and N. Warner,
Nucl. Phys. B324 (1989) 427.}: There is a natural way to define
the product of these operators which yields
a closed truncated operator algebra when restricted
to this special finite subspace of operators which becomes finite
and related to the cohomology ring\foot{This is unlike the ordinary cohomology
ring of manifold, in that the actual product of harmonic representatives
does not form a closed operator algebra.}.
There is one important subtlety however:  Unlike the
ordinary cohomology ring, the ring we get {\it depends} on the
Kahler class of the metric on $M$.  Only in the limit where
we rescale the metric $g\rightarrow \lambda g$ and let $\lambda
\rightarrow \infty$ do the ring of $A_\alpha$'s become
exactly the cohomology ring of $M$.
The deviation from the classical result is due to
instanton corrections \ref\inst{M. Dine, N. Seiberg, X.G. Wen and
E. Witten, Nucl. Phys. B278 (1987) 769; B289 (1987) 319.}
(an explicit exact result for instanton correction on
$Z$ orbifold is discussed in \ref\orin{S. Hamidi and C. Vafa, Nucl.
Phys. B279 (1987) 465\semi L. Dixon, D. Friedan, E. Martinec and
S. Shenker, Nucl. Phys. B282 (1987) 13.}).
So string theory deforms the cohomology ring.
A nice description of this deformation is as follows
\ref\witd{E. Witten, Comm. Math. Phys. 118 (1988) 411;  Nucl. Phys. B340
(1990) 281.}.  In order to describe this it
is more convenient to go to the dual basis (i.e., homology).  Let
$A^\alpha$ denote the dual basis.  Each $\alpha$ can be represented
by a cycle in $M$.  In order to specify the ring, it is sufficient
to give the trilinear pairing between cycles.  The ordinary
ring is obtained by defining this pairing to be
$$< A^\alpha A^\beta A^\gamma> =\# (C^\alpha \cap C^\beta \cap C^\gamma)$$
i.e., the number of common intersection points of the three cycles (and
defining
it to be zero if the common intersection has dimension bigger than zero).
To define the ring we obtain in string theory we have to consider
the space of holomorphic maps from $CP^1$ to the manifold $M$
(rational curves in $M$), with
the restriction that three fixed points on $CP^1$ get mapped
to points in $C^\alpha, C^\beta$ and $C^\gamma$ respectively.
Again if the dimension of moduli of such maps
is positive they do not contribute to the cohomology ring.
The isolated ones contribute weighed by the instanton action.
Let us denote an element of
the space of such holomorphic maps by $h^{\alpha \beta
\gamma}$.  Let $U$ denote the image of the sphere under $h$.
Let $k$ denote the Kahler form on $M$.
Then the definition of the deformed ring (which is commutative
and associative as shown in \witd ) is
\eqn\defo{<A^\alpha A^\beta A^\gamma>=\sum_{h^{\alpha \beta \gamma}}
\# (C^\alpha \cap U)\cdot \# (C^\beta \cap U) \cdot \# (C^\gamma \cap U)
\ {\rm exp} -\int h^{*\alpha \beta \gamma}(k)}
Note that in the limit $k\rightarrow \infty$ only the constant holomorphic
maps survive in this sum and that gives back the ordinary definition of
intersection between cycles.  So in this way we have a {\it quantum
deformed} cohomology ring.
To actually derive \defo\ in the context
of string theory (and define it properly for multiple covers of holomorphic
maps)\foot{Recent progress from this viewpoint has been made in
\ref\asmor{P.S. Aspinwall and D.R. Morrison, {\it Topological
Field Theory and Rational Curves}, preprint, OUTP-91-32p, DUK-M-91-12.}.}
is achieved by showing the topological
nature of computation (and showing that on the cylinder it can be rephrased
as a computation in a topological sigma model \witd\ which is discussed
briefly in section 6).
Without going to much detail let me at least indicate why
its form is reasonable from what we have discussed up to this point.
As we have discussed before to compute the algebra of operators
in string theory we have to consider maps of a sphere
with three discs cut out, to the manifold
with three fixed boundary circles mapped to specific loops on the manifold.
For constant loops or loops which are `close' to being constant,
we can take the limit in which the discs shrink to points, and map
a specific point on $CP^1$ to a particular point on the manifold.  Now
the string loop amplitude computation tells us that we have
to sum over all such loops weighed with $e^{-E}$, which in this
case is nothing but the exponential of the pull back of the
Kahler form on the map, as it appears in \defo .
The factors in front of exponential simply counts how many
inequivalent ways a fixed rational curve could
map to the three cycles (which is accomplished by an $SL(2,C)$ transformation
of $CP^1$ to move the three points on the sphere).
The fact that we sum over only holomorphic maps in \defo\ and
get an exact answer and its precise definition
can be best understood in the topological
description of sigma models \witd .

It is quite natural to speculate that
this deformed ring may be the actual
cohomology ring on a properly defined loop space.
One way this may be realized is to consider the space of holomorphic maps
from the disc to the manifold.  The
map from the boundary of the disc to the manifold induced from such maps
may be viewed as a `modified' loop space.
In this loop space the points of the manifold will
be represented more than once in the loop space; in fact if we look
for the space of constant loops which was
previously  isomorphic to the manifold,
that would be the same as looking for
holomorphic maps which take the boundary of the disc to a
point, which is basically
a holomorphic map from the sphere to the manifold.  So
in this case the manifold and all the holomorphic curves in it are representing
the original manifold in this loop space.
In this set up it is likely to expect that there
exists a fixed point formula for the cohomology elements (corresponding to
the circle action on the loop) which reduces the computation of cohomology
elements to the fixed point subspace which consists of the manifold and the
holomorphic curves in it.  This would then (presumably)
give rise to the cohomology ring
defined in \defo\ with $k=0$.  We can then expect to get the deformed
ring by twisting the cohomology ring, which allows us to weigh the different
fixed points (i.e., different holomorphic maps) differently, and thus
obtain the formula \defo\ with $k\not= 0$.
 This line of thought is worth pursuing further and may lead to a better
geometrical understanding of the loop space itself.

As an example, if one considers strings on $CP^1$, if we denote by
$x$ the standard $(1,1)$ cohomology element, the classical cohomology ring
is generated by $x$ with
$$x^2=0$$
Let $\beta ={\rm exp}-\int k$ integrated
over the nontrivial 2-cycle. Then the quantum deformed cohomology ring
can be computed from its definition given above and is
generated by $x$ but the relation is deformed to \witd\
$$x^2 =\beta $$
This can be generalized to $CP^n$ \ref\Int{K. Intriligator,
{\it Fusion Residues}, Harvard preprint, HUTP-91/A041.}
with the result that the quantum cohomology ring is defined by
$$x^{n+1}=\beta$$
We will discuss the conjectured
 generalization of this to the Grassmanians in the next
section (see also \Int ).

Note that in the above examples the deformed or quantum cohomology
ring does not respect the grading of differential forms (in physics
terminology we say that the instantons have destroyed chiral fermion
number conservation), but the amount of violation of grading can be
understood. The point is that the (formal) dimension of moduli space ${\cal
M}$ of
holomorphic maps $h$ is given by
$${\rm dim}{\cal M}= d+c_1(h)$$
where $d$ is the dimension of manifold and $c_1(h)$ denotes the
evaluation of the first chern class on the image of $h$.
By the definition of quantum cohomology ring we see that the
sum of dimensions of cohomology elements will have to be
$d+c_1(h)$ in order to get a non-vanishing result, which means
that we have a violation by $c_1(h)$.  This explains the
cohomology ring structure for $CP^n$ discussed above (where
$c_1=n+1$ for the fundamental cycle).  Note the fundamental
role played by Kahler manifolds where $c_1=0$, i.e., the Calabi-Yau
manifolds.  In this case there is no violation of the grading, and we
indeed get a quantum cohomology ring which respects the cohomology
grading.  For Calabi-Yau manifolds of dimensions one and two (torus
and $K3$), there are generically no holomorphic maps
(this is due to the fact that if there were any there would be
a three dimensional family of them by Mobius transformations, and
so this would be in contradiction with the above formal dimension).
So the first case of interest in terms of the deformation of
cohomology rings is the case of Calabi-Yau 3-fold, which has
also been the case of most interest for string theory\foot{
For manifolds which have $c_1<0$, by which I mean there
are some two cycles where $c_1$ evaluates to a negative number,
the underlying theory is not very well behaved (i.e., it
is not asymptotically free) and it seems that similarly
the quantum cohomology ring is somewhat ill defined
(in the Landau-Ginzburg description to be mentioned in section 5
it corresponds to perturbing the action by non-renormalizable
terms with charge greater than 1).  So quantum cohomology
rings make better sense for $c_1\geq 0$.  However it would be interesting
to see, and there is some indication \ref\hor{This idea arose in discussions
with G. Horowitz.}\ that maybe the
mirror map acts on the space of {\it all}
Kahler manifolds (possibly non-compact)
by flipping the sign of $c_1$, which in particular
sends a Calabi-Yau manifold to
another Calabi-Yau manifold.}.

%

To obtain a `static' solution to superstring theory, it turns
out that the target Kahler manifold $M$ should admit a Ricci-flat
metric, i.e., by Yau's theorem
it should be a Calabi-Yau manifold\foot{Physically
we should not ignore other manifolds as is commonly done, since
one can use them to construct interesting non-static
solutions of string theory, of the type relevant for cosmology
(see for example \ref\cosm{A. Tseytlin and C. Vafa, preprint
HUTP-91/A049; JHU-TIPAC-910028.}}.
In such a case
the dimensions of $H^{d,0}(M)$ is one, and thus we
have in our theory an operator corresponding
to this element in $H^{d,0}$ .  This operator induces
an isomorphism on the space of operators by multiplication \lvw .
This is known as the spectral flow and gives the isomorphism
$$O_{p,q}\sim O_{p+ d,q}$$
The fact that this is an isomorphism is related to the existence of
the conjugate (or inverse) operator.  In other words By conjugation
{\it there must also exist conjugate operators} in $H^{-d,0}$.
This operator induces a correspondence between
operators:
$$O_{p,q}\sim O_{p-d,q}$$
(similar statements of course hold for conjugate sectors of the Hilbert
space and amounts to shifting the anti-holomorphic degree by $-d$).
This isomorphism in particular applies to the special operators
operators $H^{p,q}$ with $0\leq p,q\leq d$
which represent cohomology of $M$ and thus suggests that there are also
`special' operators which we denote by $H^{p-d,q}$, by
shifting the holomorphic degree by $-d$.  These special operators have
the following properties which follows by the above isomorphism:
$$\rm dim \ H^{0,0}=\rm dim \ H^{-d,0} = \rm dim \ H^{0,d}=\rm dim \
H^{-d,d}=1$$
$$\rm dim \ H^{-p,q}=\rm dim \ H^{d-p,q}= h^{d-p,q}$$
where $h^{*,*}$ denote the hodge numbers of $M$.  It looks as if the
operators in $H^{-p,q}$ describe the cohomology of a $d$-dimensional manifold
which has the same hodge diamond as $M$ except that it is flipped.
In fact from the structure of string
vacuum \lvw\ it follows that
there is a closed operator ring among these states which is additive in
terms of their $Z\oplus Z$  grading just as was the case
for the operators $H^{p,q}$ with $0\leq p,q\leq d$.  Note that the
correspondence between cohomology elements of $H^{-p,q}$ and $H^{d-p,q}$
do not respect the ring structure and is thus not an isomorphisms of these
rings.  So we learn that {\it for any Calabi-Yau manifold we find
not one but two rings--only one of which is related to the deformed
cohomology ring of the manifold}.
This second ring we call the {\it complex
ring} of the manifold as it will turn out to (generically)
characterize the complex
structure of the Calabi-Yau manifold\foot{The complex ring
can be viewed geometrically as the ring generated by wedging
$H^q(\Lambda^p \Theta )$ where $\Theta$ represents the holomorphic
tangent bundle \ref\Cec{S. Cecotti,
Int. J. Mod. Phys. A6 (1991) 1749\semi
S. Cecotti, Nucl. Phys. B355 (1991) 755.}.  That their dimension
is related to that of $H^{d-p,q}$ can be easily infered from the
existence of a holomorphic $d$ form for the Calabi-Yau case.}.

So far we have described the Hilbert space and operators corresponding
to strings imbedded in a Calabi-Yau manifold $M$.  But usually
we are given not a Calabi-Yau manifold, but the string vacuum itself, i.e.,
a Hilbert space and a set of operators acting on it.
Note that in the isomorphism class of string vacua, if we just
relabel the labels of $O_{p,q}$ by $O_{-p,q}$ we have not changed
the string vacuum, and we obtain an isomorphic vacuum.  This
involution of one of the gradings
simply exchanges the two rings that we discussed above.
   In this abstract
setting how do we decide which of these two rings are `preferred' in
the sense that it corresponds to the deformation of the cohomology ring of
a manifold?  Since these two rings are absolutely on the same footing
as far as the string vacuum is concerned, i.e., that there
is an isomorphic string vacuum which relabels the sign of one of the
gradings, the only way to restore
the impartiality is {\it to postulate that for every Calabi-Yau manifold}
$M$ {\it there is another manifold} ${\tilde M}$, {\it such that
the string vacuum on either} $M$ {\it or} $\tilde M$ {\it gives rise
to both cohomology rings}.  This in particular means that
\eqn\dih{h^{p,q}({\tilde M})=h^{d-p,q}(M)}
This is the basic idea of mirror
symmetry \ref\ld{L. Dixon, unpublished.} \lvw .  Note that
this idea applies to a Calabi-Yau manifold of any dimension (not
just three as is mostly applied to).  Also
note that the dimension of complex deformations of $M$ which
is equal to $h^{1,d-1}(M)$ is equal to the dimension of Kahler
deformations of $\tilde M$ and vice versa.  So under this
mirror symmetry the shape and size of the manifolds get
exchanged.   Since the quantum
cohomology ring encodes the information about the Kahler class in it,
and under mirror symmetry shape and size get exchanged, this explains why
the second ring, the complex
ring, is characterizing the complex structure of the manifold.

Let us consider the simplest examples of mirror pairs:  As
we have discussed before for bosonic strings, strings propagating
on a torus and the dual torus give identical vacua and form mirror pairs.
It turns out that these are in fact also the simplest
examples of mirror vacua for fermionic strings.  Let us explain
this briefly in the context of simplest complex torus, a one dimensional
complex torus which is geometrically the product of two
circles with radii $R_1$ and $R_2$.  Then the complex structure
$\tau$ of the torus and its volume $-i\rho$ are given by
$$\tau =i {R_1\over R_2} \qquad \rho =i R_1R_2$$
Now we apply the duality described in section 2 for bosonic
strings in the case of target space being a torus.  This duality
works equally well for bosonic and fermionic strings.
Let us apply that to the second circle of this
example sending $R_2\rightarrow 1/R_2$ and we thus end up exchanging $\rho
\leftrightarrow \tau$ .  This is an example of the general phenomena
described above namely that the moduli controlling the shape and
the size of the Calabi-Yau manifolds are exchanged under such a
duality\foot{ This duality extends to the full moduli of the torus
not just to the case that it is geometrically the product of two circles.
In the more general case the size also is a complex modulus
due to the appearance of the anti-symmetric tensor fields which
effectively complexifies the Kahler cone.}.
This is the simplest example of mirror symmetry.
It is worth emphasizing that the other beautiful examples
that have been found are highly
non-trivial to describe geometrically \ref\examples{
B.R. Greene and M.R. Plesser, Nucl. Phys. B338 (1990) 15}
\ref\cand{P. Candelas, X.C. de la Ossa, P.S. Green and L. Parkes,
Nucl. Phys. B359 (1991) 21; Phys. Lett. 258B (1991) 118\semi
P. Candelas, M. Lynker, and R. Schimmrigk, Nucl. Phys. B341 (1990) 383.}
\ref\alr{P.S. Aspinwall, C.A. Lutken, and G.G. Ross, Phys. Lett. 241B
(1990) 373\semi P.S. Aspinwall and C.A. Lutken, Nucl. Phys. B355 (1991)
482.}\ and have far more out reaching consequences.  Nevertheless
the basic idea remains the same, and fits very naturally into
the general framework of duality just as we saw for the bosonic strings.

\newsec{Catastrophes and Superstring Vacua}
In this section we describe a link between string vacua and catastrophe
theory.  The origin of this direction of study of strings was motivated
by trying to ignore geometry of target
space and classify all string vacua
directly (as had been emphasized by Friedan).
So far we have mostly described string vacua arising from strings propagating
in some target space.  However, there are other useful
ways to describe string vacua which may or may not be related to such a
picture.
The main idea is to note that the string amplitude was defined as a sum
over all interpolating Riemann surfaces weighed by energy functional
$exp(-E)$.  Here
$E=\int |Dx|^2$ where $x$ denotes the map which defines an immersion
of the Riemann surface into the target space
(with appropriate addition of fermionic terms in the
case of superstrings).  The basic idea
to generalize this is to think of $E$ as a functional of some
fields (functions) defined on the Riemann surface.  This defines
a quantum field theory in two dimensions.
There are many interesting examples of such field theories, but
we will mention the one most relevant for superstring vacua
which is the case of Landau-Ginzburg theories.
Without going to too much detail it turns out that in this
case the field theory is characterized by a single holomorphic
function $W(x_i)$ where $x_i$ are superfields.  It was found
\ref\Cat{C. Vafa and N. Warner, Phys. Lett. B218 (1989) 51
\semi E. Martinec, Phys. Lett B217 (1989) 431.}\
that quasi-homogeneous $W$'s which have an isolated critical
point at $x_i=0$ give rise to a nice class of (super conformal) theories.
In this way the classification of quasihomogeneous singularities
became very relevant for the classification of string vacua.  Moreover,
it was found \ref\Caly{B.R. Greene, C. Vafa and N. Warner, Nucl. Phys.
B324 (1989) 371\semi E. Martinec, {\it Criticality, Catastrophe and
Compactifications}, V.G. Knizhnik memorial volume, 1989.}\
that if the index of the singularity\foot{For a quasihomogenous
function the index is defined as follows:  By assigning degree one
to a quasi-homogeneous $W$ we can obtain fractional weights
$q_i$ of variables $x_i$.  The index of $W$ is simply $\sum (1-2q_i)$.}
 is integral and equal to the number of variables $x_i$  minus $2$,
they are related to string vacua propagating on the
Calabi-Yau `manifold' defined
by (the possibly  singular variety) $ W(x_i)=0$ in weighted projective
space with a very particular Kahler metric.  This
clarified the geometrical meaning of
the important discovery of Gepner \ref\Gepi{D. Gepner, Phys. Lett. 199B
(1987) 380; Nucl. Phys. B296 (1987) 380.}\ in his construction of
string vacua.  Note that the
complex structure of the Calabi Yau is fixed by $W=0$, but
the Kahler structure of Calabi-Yau is only implicitly specified
by $W$ (through its quantum symmetries) \ref\VQ{C. Vafa, Mod. Phys. Lett.
A4 (1989) 1615.}\Caly .  As an example if we take
$$W=x^4+y^4+z^2+a\ x^2y^2$$
Setting $W=0$ in weighted projective two space, we get
a one dimensional torus whose moduli is fixed by $a$.  The volume
of the torus is implicitly fixed (by the existence of quantum
$Z_4$ symmetry)
which teaches us that the volume of this torus is $1$ for all $a$ (and the
anti-symmetric field vanishes) \VQ .
So in this way the study of strings propagating on Calabi-Yau
manifolds can be very effectively studied using this picture,
and this has become an important tool in the recent discovery
of interesting class of examples of mirror symmetric pairs of string vacua.

For strings on Calabi-Yau manifolds, as we discussed before
we automatically get two rings, only one of which is the cohomology
ring of the manifold.  What is the other,
the complex ring, geometrically?  Well,
a {\it subring} of this second ring can be described geometrically, when the
Calabi-Yau theory is represented by a variety defined by $W=0$
in a weighted projective space.
In this case if we consider the (integral dimension) ring
of the singularity defined by
\eqn\sir{{\cal R}={C[x_i]\over dW}}
they generate a subring of $H^{-p,p}$ discussed before (where
$p$ corresponds to the degree of the ring element).  It would be
interesting to see if one can extend this picture to the full ring
for all $H^{-p,q}$ (and not just the diagonal elements).
Note that this ring certainly does depend on the complex moduli
of Calabi-Yau manifold (as that changes as we change $W$).  This
is consistent with the mirror picture, namely the mirror ring
depends on Kahler moduli (as the quantum deformed cohomology
ring does depend on Kahler moduli).

So can we describe
the quantum deformed cohomology ring of some manifolds using singularity
ring for some $W$?  The answer to this question should be in the
affirmative if the mirror picture is valid.  After all the mirror map
changes $H^{-p,p}\rightarrow H^{p,p}$ and so maps (part of) the
singularity ring to the diagonal elements of the deformed cohomology ring.
The computation of cohomology ring for Calabi-Yau manifolds is in
general rather difficult.
So in this way we map a difficult problem (computation of deformed
cohomology ring) to a simple problem (computation of the ring
of a singularity) once we know the right transformation.

The quantum cohomology rings are easy to compute in some cases,
as we mentioned before.  For example for $CP^n$ we mentioned that the
deformed cohomology ring is
$$x^{n+1}=\beta$$
This of course can be written in the `mirror' picture by the ring of
$W$ according to \sir :
$$W(x)={x^{n+2}\over n+2}-\beta \ x$$
This can also be generalized to Grassmanians\foot{The cohomology
ring of Grassmanian $U(n+k)/U(n)\times U(k)$
can be written as the singularity ring \lvw\ generated
by a single potential $W(x_i)=\sum z_i^{n+k+1}/n+k+1$ where
$x_i$ are symmetric polynomials of degree $i$ in $z_j$ (with no monomial
appearing more than once) and $i$ runs from $1$ to $n$.  The
$x_i$ correspond to the chern classes of the $n$-dimensional
tautological vector bundle on the Grassmanians.  The quantum
deformation of this ring is naturally conjectured
to be $W\rightarrow W-\beta x_1$.  The motivation for this
comes from the fact that
$c_1=n+k$.}.
As discussed before for the case where $c_1\not= 0$ we expect
to violate the grading of the ring, which means the corresponding
$W$ would not be quasi-homogeneous.  For Calabi-Yau manifolds
as we mentioned before the grading of the ring is respected by the
deformation, so if the deformed ring is that of a singularity
ring the corresponding $W$ will again be quasi-homogeneous.

\newsec{Topological Mirrors}
So far we have talked about mirror symmetry in the following sense:  We
have strings propagating on two manifolds $M_1$ and $M_2$, which lead,
as described before, to two Hilbert spaces each equipped with an infinite
set of operators acting on them.  Then if these two structures, or vacua,
are isomorphic, we call $M_1$ and $M_2$ mirror pairs.  Establishing
this isomorphism at the level of Hilbert spaces is in general
a complicated task.  It would have been nice if there were a simple
criterion to establish their equivalence.  This question is also
the same as asking how do we find a simple way to classify string vacua.

Classifying string vacua (and in particular static solutions which
correspond to conformal field theories in two dimensions) has been
investigated intensively in the past seven years.  We are unfortunately
still far from a complete classification.  However for the
fermionic vacua, an interesting class of vacua have,
as discussed before, a simple description in terms of quasi-homogeneous
singularities.  In fact it is believed that for any quasi-homogeneous
function $W$ there is a unique string vacuum.  In other words it
is believed that the information about $W$ is enough to reconstruct
the full Hilbert space of strings and operators acting on it.
More generally, whether or not the theory comes from a quasi-homogeneous
singularity, it is believed that essentially given the chiral rings
in the theory one has enough
information to reconstruct the full theory.
Applied to the special case of strings propagating on manifolds
this may sound a little strange:  We seem to be
saying that given the cohomology ring of a manifold, we can find the
manifold, which is certainly false.  However it is
for the special case of Calabi-Yau manifolds that we are
considering this and in such cases just specifying the hodge
numbers may go a long way in determining the manifold itself.  Moreover
we have {\it two} rings
the {\it quantum
cohomology ring} and the {\it complex ring}, which fix the Kahler class
and the complex structure of the manifold respectively.
Thus from Yau's proof of Calabi's conjecture which
shows that knowing the Kahler class uniquely fixes the
Ricci flat metric we can reconstruct the metric on the manifold by
the information encoded in these rings.

Having said all these, it becomes clear that the phenomena of mirror
symmetry can be formulated more compactly by stating that the
two rings we get for one manifold are exchanged in the mirror manifold.
In other words we can forget about the rest of the structure of string
vacua and Hilbert spaces and the full set of operators acting on them
and concentrate simply on this finite dimensional subset of special
operators. In fact this concept can be formalized.  Consider strings
propagating on a Kahler manifold.  It turns out there is a {\it twisted}
or {\it topological} version of this theory \witd
\ref\Eg{T. Eguchi and S.-K. Yang, Mod. Phys. Lett. A4 (1990) 1653.}\
which can be obtained
by a simple modification of the definition of the theory (by
shifting the spin of fermions)
which has the effect that the only physical operators we obtain
are the ones corresponding to the cohomology classes
and that they form the quantum cohomology ring of the manifold.
If in addition the manifold in question is a Calabi-Yau manifold
this twisting can be done in two inequivalent ways (
by shifting the spins of fermions chirally, which is allowed for
Calabi-Yau manifolds because of absence of sigma model anomalies since
$c_1(M)=0$), one of which gives the quantum cohomology ring and the
other gives the complex ring, which (except for the diagonal elements) has
a less clear geometrical meaning.  In this way we can get both
rings depending on which twist we choose.  However, it is clear that
in this topological description the ordinary cohomology ring has a more
`natural' origin, and it seems to be `preferred'.
However, there is another way to describe (fermionic) string vacua and
that is via a  Landau-Ginzburg theory.
  In this case we can also twist the
theory and obtain a topological version \ref\toplg{C. Vafa, Mod. Phys.
Lett. A6 (1991) 337.}\ whose
only (physical) operators correspond to the singularity ring of $W$.
Again, if $W$ is quasi-homogeneous, this can be done in two different
ways, one of which corresponds to the singularity ring
which when $W$ describes a Calabi-Yau manifold
correspond to its complex ring, and the other
which has a less clear geometrical meaning (as it appears in the twisted
sectors) correspond to the deformed cohomology ring.  So
we see that again we have two rings, but the complex ring is
`preferred'.

The notion of mirror symmetry can be simply
translated to the {\it equivalence of a topological sigma
model with a topological Landau-Ginzburg model}, where the
`preferred' ring of the sigma model (the quantum cohomology ring
or Kahler ring)
gets mapped to the `preferred' ring of the Landau-Ginzburg model
(the singularity ring or complex ring).  Stated in this way this
mirror symmetry is more general than Calabi-Yau manifolds, as
the $W$ may or may not correspond to a Calabi-Yau manifold (even
if it is quasi-homogeneous (see next section)).  Also $W$ may not
be quasi-homogeneous as
the example of the Grassmannians mentioned before illustrates
(i.e., it goes  beyond conformal theories) but nevertheless we have
a mirror symmetry in the sense defined above.

\newsec{Some Puzzles and Conclusion}

It would be nice to be able to state the mirror symmetry in
full generality.
In geometrical terms, in the sense that strings on manifold $M_1$
behave the same way as strings on manifold $M_2$ this would be rather
difficult to do.
It is difficult even to fix precisely which category of geometrical objects
we are considering.  If we fix the category to be that of Calabi-Yau
manifolds this would be false because there are examples of Calabi-Yau
manifolds which are rigid (i.e., do not admit complex  deformations)
therefore their mirror would not admit Kahler deformations (i.e.,
$h^{1,1}=0$), which means that the mirror would not even be a
Kahler manifold!  So in this sense we have lost the mirror.  However
in the sense of equivalence of two topological theories, i.e.,
equivalence of a topological theory based on a sigma model and that
on a Landau-Ginzburg model this may still be possible.  In fact now we
will give an example where this is indeed what happens.
Consider a three-fold Calabi-Yau manifold
defined by taking the product of three two dimensional tori, with $Z_3$
symmetry, and modding out by a $Z_3\times Z_3$
symmetry generated by the elements
$(\omega ,\omega ^{-1},1),(1,\omega ,\omega^{-1})$
acting on the three tori, where $\omega$ denotes the $Z_3$ action.
It is possible
to resolve the fixed point singularities and obtain a smooth Calabi-Yau
manifold.  This manifold is rigid, in that it does not admit any
complex deformations $h^{1,2}=0$.  The dimension of Kahler deformations is
$h^{1,1}=84$.
  What is the mirror for this manifold?  The
answer turns out to be easy in this case:  It is the Landau-Ginzburg
theory defined by
$$W=\sum_{i=1,..,9} x_i^3 +\sum_{i\not= j\not= k} a_{ijk}\ x_ix_jx_k$$
The way we know this is that at $a_{ijk}=0$ we can explicitly construct
the Landau-Ginzburg theory and compare it explicitly
with the geometrical description which also turns out to be
exactly solvable (before blowing up the singularities) and one
finds that ( with the metric and the antisymmetric field
of tori corresponding to the point
of enhanced $Z_3$ symmetry)
they agree.  Moreover
one can map the fields $x_ix_jx_k$ to the Kahler classes of the
manifold.  So in this way the $84$ Kahler deformations of the manifold
(which includes the blow up modes) will get mapped to the deformation
of $W$ which are captured through varying $a_{ijk}$ above.  This
description of mirror symmetry is enough to capture the counting
of instantons on the original manifold by studying variations
of Hodge structure characterized by $W$ \Cec
so for the purposes of `simplifying' the instanton counting it works
as well.  So in a sense we do not really need a geometrical mirror;
or if we insist we can say that the geometrical mirror in this case
is a $7$ fold defined by $W=0$ in $CP^8$.  But this description
is only valid as far as we are relating the variation of its
Hodge structure with the deformed cohomology ring of the original
Calabi-Yau manifold\foot{It would be interesting to see if turning on
(possibly singular) dilaton fields and torsion on this 7-fold, gives a sigma
model which is equivalent to the three fold Calabi-Yau we started with.
As is well known turning on dilaton field shifts the effective dimension
(central charge) of the theory.
In such a picture the freezing of Kahler degrees of freedom would
be related to solving dilaton equations of motion.}.
This example reinforces another view of mirror
symmetry, namely, the abstract property of the rings that may arise
in conformal theory is the same whether or not they come from the
cohomology ring or the complex ring.  So somehow the lesson is
to {\it forget} about the underlying manifold altogether and concentrate
on abstract properties of the rings, and the classification of
the kinds of rings that can appear.  This is very much the question of
classification of variation of Hodge structures\ref\gri{
P. Griffiths, {\it Variation of Hodge Structure}, in
Topics in Transcendental Algebraic Geometry, Princeton
University Press, Princeton 1984.}.  This is in fact the
point of view advocated by Cecotti \Cec .
In this setup the existence of mirror symmetry is probably
related to the `scarcity' of inequivalent types of variations of hodge
structure (with some given topological invariants).

The same idea of mirror picture applies even to the general case of
manifolds with $c_1\not= 0$, for example the Grassmanians,
where the `mirror symmetry' allows us
to compute exactly instanton corrections to the analog of `Weil-Petersson'
metric for such manifolds \ref\CV{S. Cecotti and C. Vafa, {\it
Topological Anti-Topological Fusion}, preprint HUTP-91/A031;
SISSA-69/91/EP; and HUTP-91/A062 .}.
This follows from the structure
of special geometry which exists even off criticality.  This reinforces
the picture that {\it we should not restrict our attention to Calabi-Yau
manifolds if we are to have a deeper understanding of mirror symmetry}.

The notion of `quantum' cohomology ring might remind one of seemingly
 unrelated subject of `quantum' groups.
As is well known these groups have representation ring which is a
`quantum deformation' of the classical representation ring of the group.
The deformations being  parametrized by a parameter $k$ which is
the level of quantum group, and as $k\rightarrow \infty $ we recover
the classical representation ring.  Indeed this $k$ seems to play
a very similar role to the role kahler class
$k$ plays in quantum cohomology ring in the infinite limit of which
one recovers the classical cohomology ring.
 It turns out these two different `quantum rings'
are not as unrelated as might seem at first sight!  In particular
it has been shown \ref\gepp{D. Gepner, {\it Fusion Rings and Geometry},
preprint, NSF-ITP-90-184.}\ that for special
class of such theories the fusion ring (representation ring)
of quantum groups get mapped to the chiral ring of a Landau-Ginzburg
theory (see also \ref\sps{M. Spiegelglas and S. Yankielowicz,
in preparation\semi M. Spiegelglas, {\it Setting Fusion rings in
topological Landau-Ginzburg theories}, preprint, Technion-PH-8-91 (1991).}
\Int ).
For example, if one considers $SU(n)$ quantum group with level $k=1$,
its representation ring is isomorphic
to the quantum cohomology ring of $CP^{n-1}$
(discussed before).  It would be interesting to see whether or not all
the rings of quantum groups can be interpreted as the quantum cohomology
ring of some manifold (for example  which manifold has the quantum
cohomolgy ring related to Chebychev polynomial?).
This connection has become even more intriguing with the discovery \CV\
that precisely these Landau-Ginzburg theories seem integrable field
theories in the sense that they have an integrable classical equation
describing the generalized special geometry
(some further evidence for their integrability has been found in \ref\NW{
D. Nemeschansky and N.P. Warner, {\it Topological Matter, Integrable Models
and Fusion Rings}, preprint USC-91/031.}).
It was further conjectured in \CV\ that whenever the ring
of a supersymmetric theory corresponds to that of a RCFT (rational
conformal field theories), i.e. a solution to quantum group representation
ring, the corresponding field theory is integrable.
These connections we believe are very important to understand better
for a more abstract  understanding of `mirror symmetry' and `quantum
rings'.

We have learned that mirror symmetry is the statement
of equivalence of two topological theories, one which is difficult
to compute and the other which is easy.  It is natural to continue
this line of thinking and suggest that the same thing happens for
other topological theories.  In particular Donaldson theory
which captures some invariants for differentiable manifolds in four
dimensions, has a topological field theory
description \ref\witdon{E. Witten, Comm. Math. Phys. 117 (1988) 353.}.
It is in general very difficult to compute Donaldson invariants, just
as it is in general difficult to compute the number of rational
(holomorphic) curves in a manifold.  But we have seen in the latter case
that there is a simpler topological theory which is the Landau-Ginzburg
description.  It is tempting to conjecture that there is
a similar thing going to happen in four dimensions \ref\ccv{
S. Cecotti and C. Vafa, Work in progress.}, namely that there
must be a topological mirror theory, far simpler than Donaldson theory, which
via an appropriate mirror map allows us to effectively compute
Donaldson invariants. It remains to be seen if this conjecture is valid.

It is a pleasure to thank S. Cecotti for many discussions
which has greatly influenced my thinking on this subject.
I would like to thank D. Kazhdan for a careful reading of this
manuscript and for making suggestions for its improvement.
I also am thankful to I. Singer and S.-T. Yau for encouraging
me to participate in this conference.
This work was supported in part by the Packard Foundation and NSF
grants PHY-89-57162 and PHY-87-14654.

\listrefs

\end